\documentclass[aps, prl, bibnotes, preprintnumbers, twocolumn, 10pt]{revtex4-2}
\usepackage{style}

\begin{document}

\title{Piezoelectric Bulk Acoustic Resonators For Dark Photon Detection}

\author{Tanner Trickle}
\thanks{\href{ttrickle@fnal.gov}{ttrickle@fnal.gov}, ORCID: 0000-0003-1371-4988~\orcidlink{0000-0003-1371-4988}}
\affiliation{Fermi National Accelerator Laboratory, Batavia, Illinois 60510}

\preprint{FERMILAB-PUB-25-0003-T}
\date{\today}

\begin{abstract}
    The kinetically mixed dark photon is a simple, testable dark matter candidate with strong theoretical motivation. Detecting the feeble electric field dark photon dark matter produces requires extremely sensitive detectors. Bulk acoustic resonators (BARs), with their exceptionally high-quality phonon modes, are capable of achieving incredible sensitivity to gravitational waves in the MHz to GHz frequency range. The BAR phonons are typically read out by detecting the electric field generated by the BAR materials' piezoelectricity. Here we show that this piezoelectricity also rewards such detectors sensitivity to dark photon dark matter, as the dark electric field can resonantly excite BAR phonons. A single 10\,g piezoelectric BAR in a large, cold, environment can be orders of magnitude more sensitive to the kinetic mixing parameter than any current experiment, with only a month-long exposure and thermally-limited backgrounds.
\end{abstract}

\maketitle

Ultralight bosons, with sub-eV mass, are particularly compelling dark matter (DM) candidates. They can be produced by a plethora of cosmological mechanisms, and naturally arise in many Standard Model extensions, e.g., pseudoscalar axions are a promising solution to the Strong CP problem~\cite{Weinberg:1975ui,Peccei:1977ur,Peccei:1977hh,Wilczek:1977pj}, and massive vectors appear when spontaneously breaking gauge groups. The diversity of couplings ultralight bosons can have with the Standard Model necessitates using a range of experiments to search for them. For example, cavity haloscopes (e.g., ADMX~\cite{ADMX:2018gho}), are sensitive to the electromagnetic fields generated by an axion passing through an external magnetic field~\cite{Sikivie:1983ip}, and single-phonon based direct detection experiments (e.g., TESSERACT~\cite{Chang2020}), are sensitive to phonons created by the absorption of ultralight DM~\cite{Knapen:2017ekk,Knapen:2021bwg,Mitridate:2023izi,Linehan:2024btp,Bloch:2024qqo}. Sensitive detectors primarily used for other physics purposes have also synergized with the effort to search for ultralight bosonic DM. For example, Weber bars~\cite{Weber:1960zz} and resonant mass detectors~\cite{Aguiar:2010kn}, primarily used for kHz frequency gravitational wave (GW) detection, have been shown to sensitive to scalar DM which can oscillate fundamental constants and generate strain~\cite{Arvanitaki:2015iga,Manley:2019vxy}. For recent reviews of ultralight DM, see Refs.~\cite{Antypas:2022asj,Berlin:2024pzi}.

Here we focus on a specific spin-1 ultralight DM candidate, the kinetically-mixed dark photon, whose interaction Lagrangian is given by, 
\begin{align}
    \mathcal{L} \supset - \frac{1}{4} V_{\mu \nu} V^{\mu \nu} + \frac{m_V^2}{2} V^\mu V_\mu - \frac{\kappa}{2} \, V_{\mu \nu} F^{\mu \nu} \, ,
    \label{eq:dark_photon_lagrangian}
\end{align}
where $V_\mu$ is the dark photon field, $m_V$ is the dark photon mass, $\kappa$ is the kinetic mixing parameter, $V_{\mu \nu} = \partial_\mu V_\nu - \partial_\nu V_\mu$, and $F^{\mu \nu}$ is the electromagnetic field strength tensor. The relic abundance of dark photon DM can be generated cosmologically by a variety of different mechanisms, including the misalignment mechanism (with additional non-minimal gravitational couplings)~\cite{Arias:2012az}, inflationary production from quantum mechanical fluctuations~\cite{Graham:2015rva,Kolb:2020fwh}, topological defect decay~\cite{Long:2019lwl} or via resonances with additional particles~\cite{Bastero-Gil:2018uel,Dror:2018pdh,Agrawal:2018vin,Co:2018lka}. However it has recently been shown that if the dark photon mass is generated via a Higgs mechanism, defect production can spoil the aforementioned production mechanisms~\cite{Cyncynates:2023zwj}; further model-building can be done to alleviate these strong constraints~\cite{Cyncynates:2024yxm}.

In this \textit{Letter} we show that cm-scale piezoelectric bulk acoustic resonators (BARs) can search for dark photon DM beyond the reach of any current experiment. Their piezoelectric nature allows the dark photon electric field to resonantly drive excitations of BAR phonons, 
which can possess exceptionally high quality factors, attaining values as large as $10^{10}$~\cite{Galliou_2013}. Such piezoelectric BARs have been used previously to search for scalar DM~\cite{Arvanitaki:2015iga,Manley:2019vxy} and high-frequency GWs~\cite{Goryachev:2014yra,Goryachev:2021zzn,Campbell:2023qbf}. A single $10 \, \text{g}$ piezoelectric BAR placed in large, shielded environment, e.g., the Colossus dilution refrigerator under construction at Fermilab~\cite{Hollister:2024plk}, can achieve orders of magnitude better sensitivity to the kinetic mixing parameter than current cavity-based searches, with only a month of exposure time.

We begin with a discussion of the phonons modes inside BAR devices, illustrating how the boundary conditions can localize phonon mode profiles. We then derive the signal power delivered to a piezoelectric BAR due to dark photon DM, and assess the overall sensitivity to $\kappa$ with a variety of choices for the BAR dimensions and experimental configurations. Throughout we work in natural units where $c = \hbar = k_\text{B} = 1$.

\begin{figure}
    \centering
    \includegraphics[width=\linewidth]{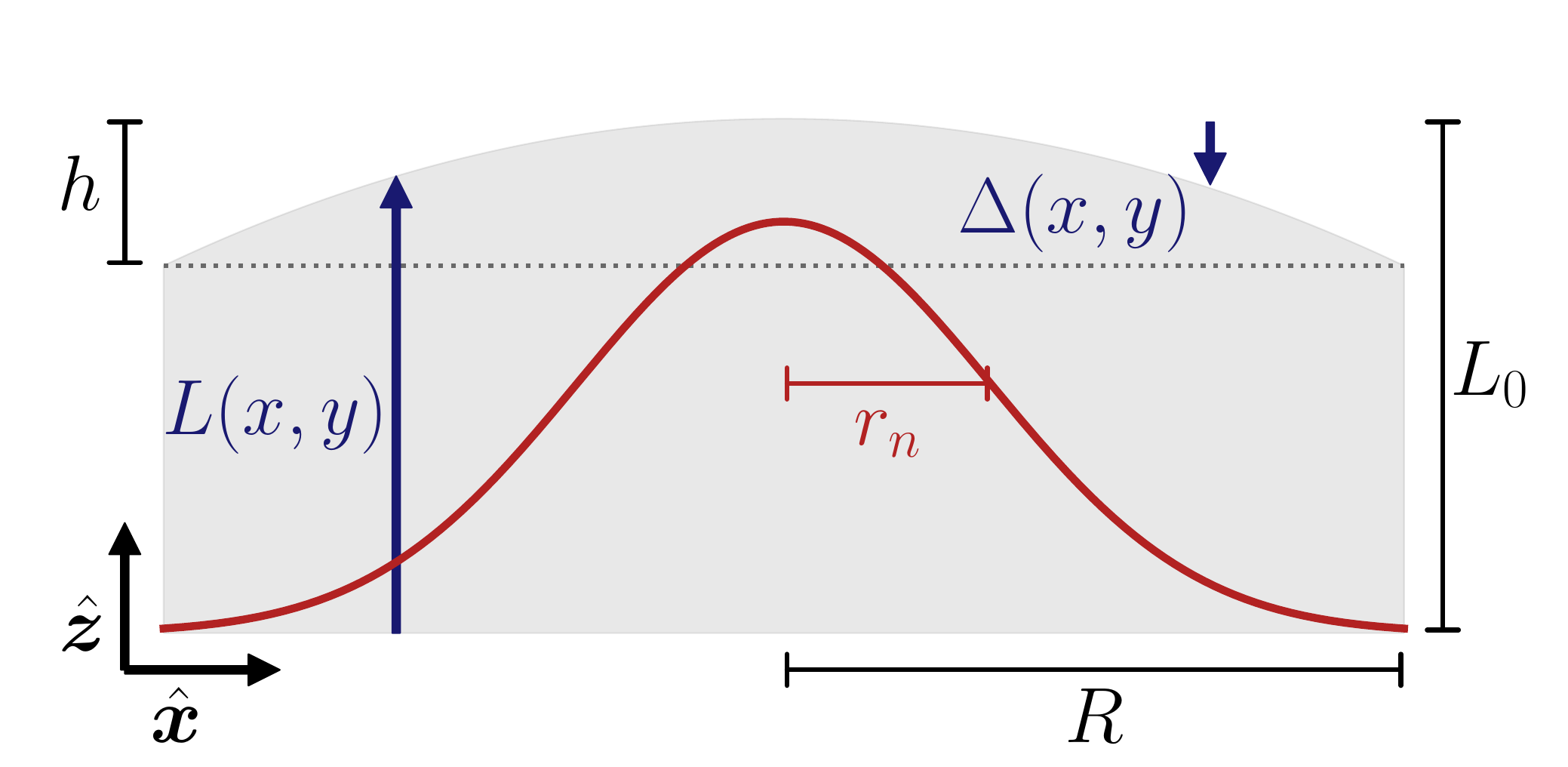}
    \caption{Cross-section schematic of a plano-convex bulk acoustic resonator. The device dimensions, $h, R, L_0$ are outlined in black, and the $x,y$-dependent distances ($L(x, y)$ and $\Delta(x, y)$) are highlighted in blue. An example Gaussian phonon mode profile is shown in red, with a radius $r_n$ given in Eq.~\eqref{eq:r_n}.}
    \label{fig:diagram}
\end{figure}

\vspace{0.25cm}
\noindent
\textbf{Bulk Acoustic Resonator Phonons.}\, A BAR is a crystal fabricated to host acoustic phonons with large quality factors ($Q_\text{p}$). Since the longest wavelength (lowest frequency) acoustic phonons inside a BAR are determined by the device dimensions, their profiles can be engineered by changing the BAR geometry. A common choice for the BAR shape is a plano-convex geometry, or a cylinder whose top is an inverted parabola~\cite{Goryachev:2014yra}. This geometry admits phonon modes with Gaussian profiles which taper towards the edge; a crucial feature to exponentially avoid losses from the BAR edge~\cite{kharel2018ultrahighqphononicresonatorsonchip}. A cross section of the plano-convex BAR we consider here is shown in Fig.~\ref{fig:diagram}. $L(x, y) = L_0 - \Delta(x, y)$ is the length in the $\hat{\vec{z}}$ direction and $L_0$ is the length in the center ($x = 0, y = 0$). $\Delta(x, y) = h \, ( x^2 + y^2 ) / R^2$, is the height profile of the inverted parabolic top, $h$ is the ``dip" height, and $R$ is the BAR radius. The typical hierarchy of length scales is $h \ll L_0 \ll R$.

The phonons that can occupy the BAR are those which satisfy the elastic wave equation and boundary conditions. We assume stress-free boundary conditions on the surface, although because the phonon mode profiles will be exponentially suppressed in the $\hat{\vec{x}}, \hat{\vec{y}}$ directions the stress-free boundary criteria is only important for the boundaries in the $\hat{\vec{z}}$ direction. The phonons can be read out without imposing stress in the $\hat{\vec{z}}$ direction due to the BAR piezoelectricity. Phonons in the BAR will generate an electric field that can be sensed by disconnected electrodes spatially separated from the BAR in the $\hat{\vec{z}}$ direction, as in Ref.~\cite{Goryachev:2014yra}. This requires that the electric field generated is in the $\hat{\vec{z}}$ direction. Different piezoelectrics will have different phonon modes that can generate electric fields in the $\hat{\vec{z}}$ direction. For simplicity, we focus on a crystal whose generated electric field in the $\hat{\vec{z}}$ direction is dominantly due to a displacement in the $\hat{\vec{z}}$ direction, i.e., $\vec{u} \approx u \, \hat{\vec{z}}$. Correspondingly, this means that the detector is directional, and most sensitive to the $\hat{\vec{z}}$ component of the dark electric field. 

A derivation of the phonon mode profiles, $U_n(\vec{x})$, and the necessary approximations to solve for them, are discussed in detail in the \textit{Supplemental Material}, and analogous derivations can be found in Refs.~\cite{Goryachev:2014yra,hBAR_masters,Linehan:2024btp}. The relevant mode profiles are given by,
\begin{align}
    U_n(\vec{x}) = \frac{1}{\sqrt{M_n \omega_n}} \cos{\left( \frac{n \pi z}{L(x, y)} \right)} \exp\left( - \frac{x^2 + y^2}{2 \, r_n^2} \right) \, ,
    \label{eq:phonon_mode_functions}
\end{align}
where $n \geq 1$, $M_n = \rho \, \pi r_n^2 L_0$, $\rho$ is the BAR mass density, $\omega_n$ is the phonon energy, and $U_n$ is normalized to $\int U_n^2 \, \dd^3\vec{x} = (2 \rho \, \omega_n)^{-1}$. The displacement operator is quantized in terms of these mode profiles as, $u(\vec{x}, t) = \sum_{n} U_n(\vec{x}) \left[ b_n e^{- i \omega_n t} + b_n^\dagger e^{ i \omega_n t} \right]$, where $b_n^\dagger, b_n$ are the raising and lowering operators, respectively, which satisfy the canonical commutation relations, $[b_n, b_{n'}^\dagger] = \delta_{nn'}$. The frequency of the $n^\text{th}$ mode is,
\begin{align}
    \frac{\omega_n}{2 \pi} = \frac{n c_\text{l}}{2 L_0} \sim 5 \, \text{MHz} \times n \, \left( \frac{c_\text{l}}{10 \, \text{km} / \text{s}} \right) \left( \frac{1 \, \text{mm}}{L_0} \right) \, ,
\end{align}
where $c_\text{l}$ is the longitudinal speed of sound. The Gaussian phonon mode profile radius, $r_n$, is,
\begin{align}
    \frac{r_n}{R} & = \left( \frac{c_\text{t}}{\omega_n R} \right)^{1/2} \left( \frac{L_0}{2 h} \right)^{1/4} \nonumber \\
    & \sim \frac{10^{-1}}{n^{1/2}} \left( \frac{L_0}{1 \, \text{mm}} \right)^{3/4} \left( \frac{10 \, \text{cm}}{R} \right)^{1/2} \left( \frac{10 \, \mu \text{m}}{h} \right)^{1/4} \, ,
    \label{eq:r_n}
\end{align}
where $c_\text{t}$ is the transverse sound speed, and we have assumed $c_\text{t} \approx c_\text{l}$ in the parametric expression. There are two competing factors in the optimization of $r_n$ for dark photon detection. $r_n$ must be small to minimize losses through the edges (and therefore achieve a large $Q_\text{p}$), but it must also be large to increase the effective mass of the detector, $M_n$, which increases the coupling to dark photons, which we will now discuss in detail.

\setlength{\tabcolsep}{7pt}
\begin{table*}[ht!]
    \begin{center}
        \begin{tabular}{lcccccc} \toprule
         \multirow{2}{*}{\textbf{Design}} & \multicolumn{4}{c}{\textbf{Resonator}} & \multicolumn{2}{c}{\textbf{Experiment}} \\ 
         \cmidrule(lr){2-5} \cmidrule(lr){6-7}
         & Length ($L_0$) & Dip height ($h$) & Radius ($R$) & Mass ($M$) & Shield Size ($R_\text{s}$) & Temperature ($T$) \\\midrule
         \textcolor{ForestGreen}{MAGE} & 1 mm & $0.5$ mm & 15 mm & 2 g & 10 cm & 4 K \\
         \textcolor{LimeGreen}{MAGE - cold} & 1 mm & $0.5$ mm & 15 mm & 2 g & 10 cm & 10 mK \\\midrule
         \textcolor{ProcessBlue}{Colossus - Broad} & 1 cm & 44 $\mu$m & 15 cm & 2 kg & 1 m & 20 mK \\
         \textcolor{NavyBlue}{Colossus - Peak} & 62 $\mu$m & 10 nm & 15 cm & 10 g & 1 m & 20 mK \\\midrule
         \textcolor{BurntOrange}{Cryo Tank - Broad} & 1 cm & 44 $\mu$m & 15 cm & 2 kg & 10 m & 4 K \\
         \textcolor{BrickRed}{Cryo Tank - Peak} & 620 $\mu$m & 11 nm & 15 cm & 110 g & 10 m & 4 K \\\bottomrule
        \end{tabular}
    \end{center}
    \vspace{-1.5em}
    \caption{Summary of the experimental parameters assumed for each design. ``Resonator" parameters define the physical dimensions, and mass $M \approx \rho \, \pi R^2 L_0$, of the BAR. ``Experiment" parameters define the size of the shielding environment, $R_\text{s}$, and the physical temperature of the BAR, $T$. Experimental parameters for the MAGE designs are from Ref.~\cite{Campbell:2023qbf}, and the parameters of the ``Colossus" dilution refrigerator are from Ref.~\cite{Hollister:2024plk}.}
    \vspace{-1.5em}
    \label{tab:designs}
\end{table*}

\vspace{0.25cm}
\noindent
\textbf{Signal.}\, The dark photon interaction in Eq.~\eqref{eq:dark_photon_lagrangian} will generate an effective, ``dark", electric field, $\vec{E}'$, which couples to phonons in the piezoelectric BAR. This can be understood as a direct mixing between the photon and dark photon, as in Eq.~\eqref{eq:dark_photon_lagrangian}, or by transforming to the mass basis, $A_\mu \rightarrow A_\mu - \kappa V_\mu$. In the mass basis $V_\mu$ couples to $U(1)_\text{EM}$ charged fields, $\psi$, as $\kappa Q V_\mu \bar{\psi} \gamma^\mu \psi$, where $Q$ is their electromagnetic charge, and therefore $\kappa V_\mu$ acts as an effective electromagnetic potential. In the long-wavelength limit the dark electric field generated is dominated by the time derivative of the vector potential, and in free space is, $\vec{E}' \approx \kappa \partial_t \vec{V} \approx \kappa \sqrt{2 \rho_V} \vec{\epsilon}_V \cos{(m_V t)}$, where $\rho_V \approx 0.4 \, \text{GeV} / \text{cm}^3$ is the local DM density, and $\vec{\epsilon}_V$ is the dark photon polarization. The dark photon can be considered long-wavelength since its de Broglie is much larger than the experiment, $\lambda_V = 2 \pi / (m_V v) \sim 100 \, \text{m}  \, \left( \mu\text{eV} / m_V \right) $, where $v \sim 10^{-3}$ is the typical local DM velocity.

There are two effects which suppress the dark electric field inside the BAR relative to free space: screening and shielding. Since the BAR is also dielectric the dark electric field is suppressed, $\vec{E}' \approx \kappa \sqrt{2 \rho_V} \vec{\epsilon}_V \cos{(m_V t)} / \varepsilon_0$, where $\varepsilon_0$ is the low-frequency dielectric constant. The more subtle suppression is due to any conductive walls around the BAR. While these are necessary to shield the experiment from environmental electric fields, they also impose conditions on the electric fields which can exist inside the shield. While the exact suppression depends on the shield geometry, parametrically, the shielding suppression is~\cite{Chaudhuri:2014dla},
\begin{align}
    \vec{E}' \approx \frac{\kappa \sqrt{2 \rho_V} \vec{\epsilon}_V }{\varepsilon_0} \cos{(m_V t)} \, \text{min}\{ 1, (m_V R_\text{s})^2 \} \, ,
    \label{eq:dark_electric_field}
\end{align}
where $R_\text{s}$ is the size of the shield. Therefore one needs $m_V \gg 1 / R_\text{s} \sim 10^{-1} \, \mu\text{eV} \, \left( 1\, \text{m} / R_\text{s} \right)$ to be unaffected by shielding.

The dark electric field interacts with the BAR via the interaction Hamiltonian, $\delta H = - \int \vec{E}' \cdot \vec{P} \, \dd^3 \vec{x}$, where $\vec{P}^i = e^{ijk}_\text{pt} \, \nabla^j \vec{u}^k$ is the polarization vector expressed in terms of the piezoelectric coefficients, $e_\text{pt}^{ijk}$, and displacement operator. Given the device geometry in Fig.~\ref{fig:diagram}, with $L_0 \ll R$, the polarization vector will be dominated by the gradient in the $\hat{\vec{z}}$ direction. Additionally, our focus is on targets which dominantly couple the $\hat{\vec{z}}$ components of the displacement and electric fields, i.e., $e_\text{pt}^{zzz} \gg e_\text{pt}^{i z k}$, for $i, k$ not equal to $z$. In this limit the interaction Hamiltonian is,
\begin{align}
    \delta H \approx -\kappa \sqrt{2 \rho_V} \frac{e_\text{pt}}{\varepsilon_0} \cos{\theta_V} \cos{(m_V t)} \, \int \left[ \nabla_z \, u \right] \,  \dd^3 \vec{x} \, ,
    \label{eq:interaction_hamiltonian}
\end{align}
where $\cos{\theta_V} = \vec{\epsilon}_{V} \cdot \hat{\vec{z}}$, and $e_\text{pt} \equiv e_\text{pt}^{zzz}$.

Given the interaction Hamiltonian in Eq.~\eqref{eq:interaction_hamiltonian} we use Fermi's Golden rule to compute the excitation rate, or signal power deposited to the system: $P_\text{s} = 2 \pi m_V |\langle n | \delta H_0 | 0 \rangle|^2 \delta(m_V - \omega_n)$, where $\delta H = \delta H_0 \, e^{i m_V t} + \text{h.c.}$. However, since neither the dark photon or phonon mode are perfect resonances the delta function will be smeared by the larger of the linewidths. Including this smearing, and evaluating the $\langle n | \delta H_0 | 0 \rangle$ matrix element using Eq.~\eqref{eq:interaction_hamiltonian} and $|n \rangle = b_n^\dagger |0 \rangle$, the signal power is,
\begin{align}
    P_\text{s} = 32 \, \kappa^2 \, \frac{\rho_V}{\rho^2} \frac{e_\text{pt}^2}{\varepsilon_0^2} \frac{M_n}{L_0^2} \frac{m_V^2  \gamma_n \cos^2{\theta_V}}{(m_V^2 - \omega_n^2)^2 + (\omega_n \gamma_n)^2} \, ,
    \label{eq:P_s}
\end{align}
where $\gamma_n = \omega_n / Q_\text{s}$ is the signal linewidth which has a quality factor of $Q_\text{s} = \text{min}\{ Q_\text{p}, Q_\text{DM} \}$, where $Q_\text{DM} \sim 10^6$ is the effective dark photon quality factor~\cite{Berlin:2024pzi}. Since acoustic phonons in BAR devices regularly achieve quality factors between $10^6 - 10^{10}$~\cite{Galliou_2013,Goryachev:2014yra,kharel2018ultrahighqphononicresonatorsonchip}, they are well in the limit of $Q_\text{p} \gg Q_\text{DM}$, and therefore $Q_\text{s} \sim 10^6$. On resonance, $m_V = \omega_n$, Eq.~\eqref{eq:P_s} simplifies to,
\begin{align}
    P_\text{s}^\text{res} = 32 \, \kappa^2 \, \frac{\rho_V}{\rho^2} \frac{e_\text{pt}^2}{\varepsilon_0^2}\frac{Q_\text{s}}{m_V} \frac{M_n}{L_0^2} \cos^2{\theta_V} \, .
    \label{eq:P_s_res}
\end{align}

\begin{figure*}
    \centering
    \includegraphics[width=\textwidth]{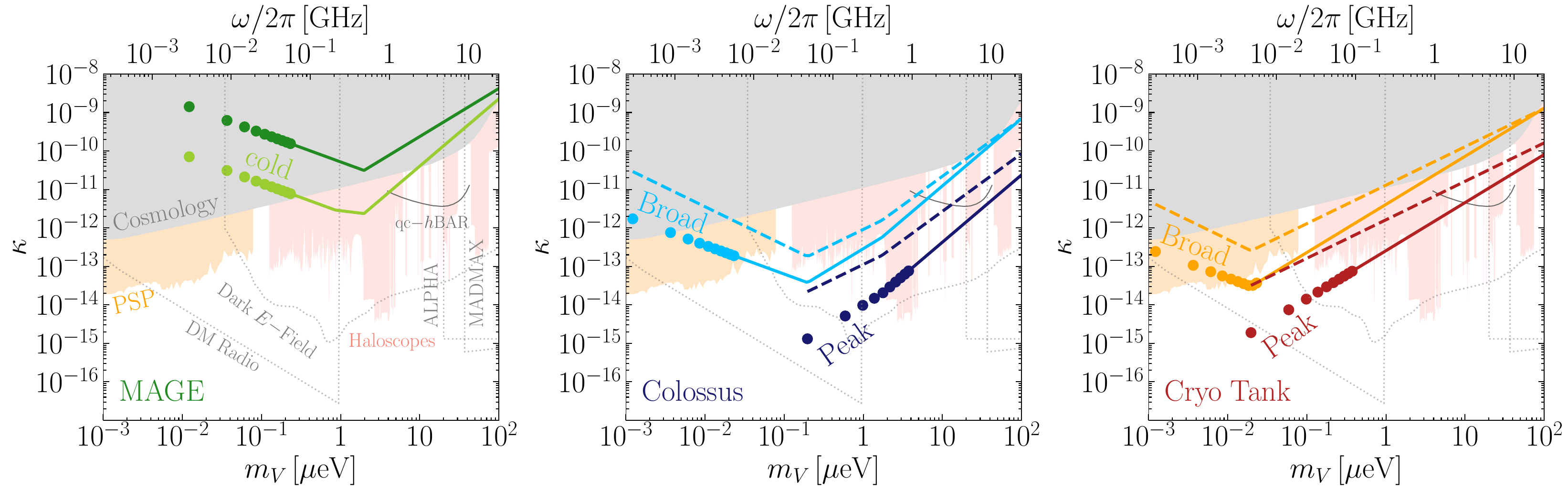}
    \vspace{-2em}
    \caption{Projected sensitivity of a single piezoelectric BAR to the kinetic mixing parameter, $\kappa$ (Eq.~\eqref{eq:dark_photon_lagrangian}), for the designs in Table~\ref{tab:designs}, averaging over the DM polarizations ($\cos{\theta_V} \rightarrow 1 / \sqrt{3}$). The sensitivity of the first ten resonances is indicated by dots; for higher modes the sensitivity lies along the extending line. Dots and solid lines assume $T_\text{obs} = 1 \, \text{month}$. Dashed lines show the scanning sensitivity, assuming $T_\text{obs} = 1 \, \text{yr}$, and simultaneous readout of all phonon modes, for each $e$-fold in $m_V$. The minimum detectable mass for each design is $m_V^\text{min} = \pi c_\text{l} / L_0$. The prominent turnover at $m_V \sim 1 / R_\text{s}$ is due to shielding effects (Eq.~\eqref{eq:dark_electric_field}). Gray shaded regions are excluded cosmologically~\cite{Arias:2012az}, orange shaded regions are excluded by recent Parker Solar Probe (PSP) bounds~\cite{An:2024wmc}, and red shaded regions are excluded by current haloscope experiments~\cite{AxionLimits}. The gray line is the projected ``Gen. II" sensitivity of a qubit-coupled high-overtone BAR (qc-$h$BAR) using single phonon readout~\cite{Linehan:2024btp}. Dotted gray lines are projections for the ALPHA~\cite{Gelmini:2020kcu}, DM-Radio~\cite{Chaudhuri:2014dla}, Dark $E$-field~\cite{Godfrey:2021tvs}, and MADMAX~\cite{MADMAX:2019pub,Gelmini:2020kcu} experiments.}
    \label{fig:sensitivity}
\end{figure*}
\vspace{-1em}

\vspace{0.25cm}
\noindent
\textbf{Sensitivity.}\, We assume that read out of the phonons generated by the signal power in Eq.~\eqref{eq:P_s_res} is done via linear amplification of the voltage generated between electrodes spatially separated in the $\hat{\vec{z}}$ direction from the BAR, as in Ref.~\cite{Goryachev:2014yra}. The signal-to-noise ratio, $\text{SNR}$, can then be determined by the Dicke radiometer equation, $\text{SNR} = \sqrt{T_\text{obs} / \Delta \nu} \, P_\text{s} / T_\text{eff}$~\cite{Dicke:1946glx}, where $T_\text{obs}$ is the observation time, $T_\text{eff}$ is the effective noise temperature, and $\Delta \nu = \omega / (2 \pi Q_\text{p})$ is the bandwidth~\footnote{The bandwidth is proportional to $1/Q_\text{p}$ versus $1/Q_\text{DM}$ since we are in the regime where $Q_\text{DM} \ll Q_\text{p}$~\cite{Cervantes:2022gtv}}. $T_\text{eff}$ is primarily determined by thermal and amplifier noise contributions~\cite{Goryachev:2014yra}. We assume that the amplifier operates at the standard quantum limit (SQL), such that its noise temperature is $\omega \approx m_V$~\cite{Caves:1982zz} when $T < \omega$, where $T$ is the physical temperature. The effective noise temperature of the system, with an amplifier that achieves the SQL, is then $T_\text{eff} = \text{max}\{ T, m_V \}$; the system is limited by thermal noise when $m_V < T$ ($T_\text{eff} = T$) and limited by quantum noise when $T < m_V$ ($T_\text{eff} = m_V$). 

The challenge of SQL readout at dilution refrigerator temperatures in the MHz to 10 GHz frequency range is shared by many ongoing axion and dark photon experiments. For frequencies below $10 \, \text{MHz}$ DC superconducting quantum interference devices (DC SQUIDs) are commonly used~\cite{Chaudhuri:2014dla}, and have shown thermally limited operation at $T = 4 \, \text{K}$, with initial measurements indicating improvements as the SQUID is cooled~\cite{Goryachev:2014nna}. In the 10 MHz to GHz frequency range readouts using AC SQUIDs have been estimated to be thermal noise limited for $T = 100 \, \text{mK}$~\cite{Mates_2008, Chaudhuri:2014dla}. The GHz and above frequency range has more options: HEMT amplifiers are commercially available but typically do not reach the SQL~\cite{Mates_2008}, and Josephson parametric amplifiers (JPAs) can achieve the SQL but have limited bandwidth~\cite{Macklin_2015} (although ``traveling-wave" JPAs can improve bandwidth~\cite{Macklin_2015}). Lastly we note that single phonon detection techniques~\cite{Linehan:2024btp} can avoid quantum noise entirely.

The specific BAR material we consider is x-cut quartz (\ce{SiO2}), whose relevant electromagnetic and mechanical parameters are given in Table~\ref{tab:parameters}. Quartz has been used extensively in BAR applications~\cite{Goryachev:2014yra,kharel2018ultrahighqphononicresonatorsonchip,Goryachev:2021zzn,Campbell:2023qbf,Yoon_2023}. The crystal cut determines the orientation of the crystal axes with respect to the BAR axes, shown in Fig.~\ref{fig:diagram}. X-cut quartz is oriented such that the crystal $\hat{\vec{x}}$ axis is parallel to the BAR $\hat{\vec{z}}$ axis. This orientation maximizes the coupling between the $\hat{\vec{z}}$ components of the dark electric field and BAR phonon since the largest piezoelectric coefficient in \ce{SiO2} is $e_\text{pt}^{xxx} \equiv e_\text{pt}$~\cite{Jain_2013,de_Jong_2015,materials_project} (which is $e_\text{pt}^{zzz}$ in BAR (Fig.~\ref{fig:diagram}) coordinates). X-cut quartz has been considered specifically as a BAR in Ref.~\cite{Yoon_2023}.

With the material parameters for x-cut quartz in Table~\ref{tab:parameters} we can now compute the expected sensitivity to the kinetic mixing parameter, $\kappa$, using the Dicke radiometer equation discussed previously. The sensitivity to $\kappa$ is given parametrically as,
\begin{align}
    \kappa & \sim 10^{-16} \,\left( \frac{m_V}{10^{-2} \, \mu\text{eV}} \right)^{5/4} \left( \frac{L_0}{1\, \text{mm}} \right)^{1/4} \left( \frac{h}{10 \, \mu\text{m}} \right)^{1/4} \nonumber\\ 
    & \; \times \left( \frac{10 \, \text{cm}}{R} \right)^{1/2} \left( \frac{10^8}{Q_\text{p}} \right)^{1/4} \left( \frac{T_\text{eff}}{10 \,\text{mK}} \right)^{1/2} \left( \frac{\text{month}}{T_\text{obs}} \right)^{1/4} \, ,
    \label{eq:kappa_sensitivity}
\end{align}
assuming an $\text{SNR} = 1$. Note that this scaling assumes no shielding suppression, which is included by multiplying by $\text{max}\{1, ( m_V R_\text{s} )^{-2} \}$, as discussed before Eq.~\eqref{eq:dark_electric_field}. The improvement to the sensitivity from $Q_\text{p}$, $T_\text{eff}$, and $T_\text{obs}$ is a direct consequence of the Dicke radiometer equation, and a larger BAR radius, $R$, simply increases the effective mass, $M_n$, since $r_n \propto R^{1/2}$ (Eq.~\eqref{eq:r_n}). The dependence of the sensitivity on $L_0$ and $h$ is a bit more subtle. While increasing $L_0$ increases $M_n$, it also increases the phonon mode wavelengths. This is detrimental to the sensitivity since the dark electric field couples to the BAR phonons via $\nabla_z u$. However, $L_0$ cannot be too small since there is another trade-off: the smallest DM mass a BAR device is sensitive to is $m_V^\text{min} = \pi c_\text{l} / L_0$. The sensitivity in Eq.~\eqref{eq:kappa_sensitivity} also improves with smaller dip heights, $h$. This is because decreasing $h$ increases $r_n$ (Eq.~\eqref{eq:r_n}), and therefore increases $M_n$. However again there is a trade-off. The smaller $h$ is, the less ``trapped" the phonon mode is, and the worse its quality factor, $Q_\text{p}$, will be. Therefore $h$ should be chosen to be large enough to ensure high-quality phonons exist, but small enough to optimize the sensitivity.

\setlength{\tabcolsep}{8pt}
\begin{table}
    \begin{center}
        \begin{tabular}{lcccc} \toprule
        \multicolumn{3}{l}{\textbf{Material: \ce{SiO2} (X-Cut Quartz)}} & \multicolumn{2}{c}{Ref.} \\\midrule
        Mass density & $\rho$ & $2.6 \, \text{g} / \text{cm}^3$ & \multicolumn{2}{c}{\cite{Jain_2013}} \\
        Static dielectric & $\varepsilon_0$ & $4.5$ & \multicolumn{2}{c}{\cite{Petousis_2017}} \\
        Piezoelectric coeff. & $e_\text{pt}$ & $0.14 \, \text{C} / \text{m}^2$
        & \multicolumn{2}{c}{\cite{Jain_2013,de_Jong_2015}} \\
        Sound speeds & $( c_\text{l}, c_\text{t} )$ & $( 5.9, 4.3 ) \, \text{km}/\text{s}$ & \multicolumn{2}{c}{\cite{Jain_2013,de_Jong_2015}} 
        \\\bottomrule
        \end{tabular}
    \end{center}
    \vspace{-1.5em}
    \caption{Relevant material parameters for x-cut quartz (\ce{SiO2}). All parameters are from the Materials Project~\cite{materials_project} for material ID \texttt{mp-7000}. Sound speeds are averaged values using the Voigt-Reuss-Hill bulk and shear moduli.}
    \vspace{-1.5em}
    \label{tab:parameters}
\end{table}

In Fig.~\ref{fig:sensitivity} we compare the sensitivity of different resonator geometries and shield configurations assuming $Q_\text{p} = 10^8$. The specific parameters for each of the designs are given in Table~\ref{tab:designs}, and the design names are color-coordinated with the projected sensitivity lines in Fig.~\ref{fig:sensitivity}. Current and projected bounds from other experiments are compiled with the help of Ref.~\cite{AxionLimits}. The projected sensitivity of the first ten resonances which couple to the dark electric field are shown as dots, and then extended as a line for visual simplicity. The dots and solid lines assume $T_\text{obs} = 1\,\text{month}$. 

Since the dark photon mass is unknown, it is also important to understand the sensitivity when the dark photon mass is scanned. In a single-mode experiment approximately $Q_\text{DM} \sim 10^6$ scan steps are required per $e$-fold in $m_V$~\footnote{An $e$-fold in $m_V$ is a range of masses whose maximum is $e$ times the minimum, e.g., $1\, \mu\text{eV} \leq m_V \leq 2.7\, \mu\text{eV}$.}, reducing the observation time per step by $Q_\text{DM}$ and requiring $T_\text{obs} \rightarrow T_\text{obs} / Q_\text{DM}$ in the Dicke radiometer equation~\footnote{We note that the observation time per step while scanning, $T_\text{obs} / Q_\text{DM}$, must also be greater than the thermalization time, $\tau_\text{th} = Q_\text{p} / \omega$~\cite{Manley:2019vxy}, which is approximately satisfied for all $\omega$ shown in Fig.~\ref{fig:sensitivity}.}. However, if multiple, $N_m$, linewidths are read out simultaneously the number of scan steps is reduced, increasing the observation time per step to $T_\text{obs} N_m / Q_\text{DM}$. The dashed lines in Fig.~\ref{fig:sensitivity} show the scanning sensitivity assuming $T_\text{obs} = 1 \, \text{yr}$, and simultaneous readout of all phonon modes, for each $e$-fold in $m_V$ ($N_m \sim m_V L_0 / (\pi c_\text{l})$~\footnote{This is the number of dominantly coupled modes which have mode functions given by Eq.~\eqref{eq:phonon_mode_functions}. Additional, more weakly coupled, modes may also prove to be useful, and are discussed in detail in the \textit{Supplemental Material}}). To realize this scanning sensitivity the BAR frequencies must be tuned. While active electromagnetic tuning has been demonstrated in Ref.~\cite{Campbell:2022jnq}, complete scanning between each resonance will likely require BARs of different $L_0$, or other forms of electric or mechanical tuning~\cite{Liu_2020,Arvanitaki:2021wjk}.

The ``MAGE" design (left panel, Fig.~\ref{fig:sensitivity}) roughly corresponds to the ongoing high-frequency GW experiment (Multi-mode Acoustic Gravitational wave Experiment (MAGE)~\cite{Campbell:2023qbf}) which uses a quartz-based BAR. A predecessor to MAGE has already been built and operated at $3.4~\text{K}$~\cite{Goryachev:2021zzn}, and used to place bounds on scalar DM~\cite{Arvanitaki:2015iga,Manley:2019vxy} and high-frequency GWs~\cite{Goryachev:2014yra,Goryachev:2021zzn,Campbell:2023qbf}. The BAR and shield dimensions in Table~\ref{tab:designs} are from Ref.~\cite{Campbell:2023qbf}. Fig.~\ref{fig:sensitivity} shows the sensitivity if readout at the SQL can be achieved over the entire frequency range. The sensitivity of the current MAGE experiment is limited to simultaneous readout of 30 modes due to readout electronics~\cite{Campbell:2023qbf}, and frequencies below $\mathcal{O}(10 \, \text{MHz})$ due to the DC SQUID~\cite{Goryachev:2014nna}. A dedicated analysis of MAGE data to search for dark photons is left for future work. ``MAGE-cold" assumes the same BAR and shield already built, but operated at dilution refrigerator temperatures, $T = 10 \, \text{mK}$. 

A main limitation in the MAGE design is its small, $R_\text{s} \sim 10 \, \text{cm}$, shield. Effectively searching for dark photon DM at sub-GHz frequencies requires a large, cold environment. The largest volume dilution refrigerator, ``Colossus"~\cite{Hollister:2024plk}, is currently being constructed at Fermilab and will have a volume of roughly $1 \, \text{m}^3$ cooled to $T = 20 \, \text{mK}$. We consider the sensitivity of two different BARs operating inside Colossus (middle panel, Fig.~\ref{fig:sensitivity}). In the first, labeled ``Peak", $L_0$ is chosen to maximize the peak sensitivity at the shielding effect boundary, $m_V \sim 1 / R_\text{s}$ ($L_0^\text{peak} \approx \pi c_\text{l} R_\text{s}$). In the second, labeled ``Broad", $L_0$ is chosen to broaden the sensitivity in mass, down to $m_V \sim 1\, \text{neV}$. In both setups $h$ is chosen such that $r_n \ll R$, which parametrically happens as long as $h \gtrsim L_0 (L_0 / R)^2$. Lastly, the ``Cryo Tank" configuration (right panel, Fig.~\ref{fig:sensitivity}) shares the same ``Peak" and ``Broad" resonator dimension optimizations as ``Colossus", but is operating inside a much larger shield, at a warmer $T = 4\, \text{K}$.

\vspace{0.25cm}
\noindent
\textbf{Discussion.}\, Phonons in bulk acoustic resonator (BAR) devices have been shown to be exceptionally sensitive probes of new physics, from scalar DM~\cite{Arvanitaki:2015iga,Manley:2019vxy} to high-frequency GWs~\cite{Goryachev:2014yra,Goryachev:2021zzn,Campbell:2023qbf}. Here we show that piezoelectric BARs, such as quartz (\ce{SiO2}), have additional sensitivity to dark photon DM. The incoming dark photon generates an electric field which, due to the BAR piezoelectricity, resonantly excites the phonons in the BAR. We have shown that a single, $\mathcal{O}(10\,\text{g})$ mass piezoelectric BAR with thermally-limited backgrounds can be orders of magnitude more sensitive to the kinetic mixing parameter, $\kappa$, than ongoing experiments with only a month of exposure (Fig.~\ref{fig:sensitivity}). This is a novel avenue to search for physics beyond the Standard Model with existing BAR experiments, such as MAGE~\cite{Campbell:2023qbf}, and inspires a new approach to dark photon DM direct detection.

Piezoelectric BARs may have uses in other, complementary, searches for physics beyond the Standard Model. In an external magnetic field of strength $B_0$ an axion will convert to an electric field~\cite{Sikivie:1983ip}. If this conversion happens within the piezoelectric BAR the signal will be similar to the dark photon DM signal considered here, and the sensitivities from Fig.~\ref{fig:sensitivity} can be rescaled~\cite{Berlin:2023ppd},
\begin{align}
    g_{a\gamma \gamma} \sim \frac{10^{-16}}{\text{GeV}} \,  \left( \frac{\kappa}{10^{-15}} \right) \left( \frac{m_a}{10^{-2} \, \mu\text{eV}} \right) \left( \frac{1 \, \text{T}}{B_0} \right) \, ,
\end{align}
where $\omega = m_a$ is the axion mass. If nuclear spin polarization can be achieved additional axion couplings can be searched for via the ``piezoaxionic" effect~\cite{Arvanitaki:2021wjk}. Furthermore, a kg-scale, cold \ce{SiO2} target shows great promise as a target for light, sub-GeV DM scattering in to higher energy, $\mathcal{O}(\text{meV})$, phonons~\cite{Griffin:2019mvc}. It is therefore worth considering if such a target can be multi-purpose, capable of simultaneously sensing single $\mathcal{O}(\text{meV})$ phonons and measuring the $\mathcal{O}(\mu\text{eV})$ phonon population, or if piezoelectric readout of down-converted higher energy phonons is possible.

\vspace{0.25cm}
\noindent 
\textbf{Acknowledgements.} \textit{I would like to thank Roni Harnik and Ryan Linehan for helpful discussions, and I am especially grateful to Asher Berlin and Tongyan Lin for providing feedback on the draft. This manuscript has been authored by Fermi Research Alliance, LLC under Contract No. DE-AC02-07CH11359 with the U.S. Department of Energy, Office of Science, Office of High Energy Physics.}

\bibliographystyle{utphys3}
\bibliography{biblio}

\appendix
\onecolumngrid

\newpage 

\renewcommand{\theequation}{S.\arabic{equation}}
\setcounter{equation}{0}

\section{\large Supplemental Material: Phonon Eigenmode Derivation}

\vspace{-0.5em}
\begin{center}
    Tanner Trickle
\end{center}

\label{app:phonon_eigenmode_derivation}

In this Supplemental Material we provide a derivation of the phonon eigenmode profiles in Eq.~\eqref{eq:phonon_mode_functions} of the Letter. We will closely follow the derivation presented in Ref.~\cite{Linehan:2024btp}; similar derivations can be found in Refs.~\cite{hBAR_masters,Goryachev:2014yra}. The phonon eigenmodes are solutions to the elastic wave equation that satisfy the boundary conditions at the edges of the bulk acoustic resonator (BAR). Our goal here is to specify the necessary simplifying approximations to gain an analytic understanding of these eigenmodes. A general analysis of these eigenmodes in an anisotropic medium requires dedicated numerical analyses, which we leave for future work. 

Our first simplifying approximation is that the BAR crystal is isotropic, such that the mechanical wave equation for the displacement, $\vec{u}$, is determined by only the transverse, $c_\text{t}$, and longitudinal, $c_\text{l}$, sound speeds, 
\begin{align}
    \frac{ \partial^2 \vec{u} }{\partial t^2} \approx c_\text{l}^2 \, \nabla (\nabla \cdot \vec{u}) -c_\text{t}^2 \, \nabla \times \nabla \times \vec{u} \, .
    \label{eq:elastic_wave_equation_isotropic}
\end{align}
Even in the isotropic limit the wave equation couples the components of $\vec{u}$. However solutions where $\vec{u}$ dominantly oscillates in a given direction exist in approximation. For example, $\vec{u} \approx u \, \hat{\vec{z}}$ is a solution when $u$ varies slowly in $\hat{\vec{x}}, \hat{\vec{y}}$: $\partial u / \partial x \sim \partial u / \partial y \ll \partial u / \partial z$. In this limit Eq.~\eqref{eq:elastic_wave_equation_isotropic} admits solutions where the $\hat{\vec{x}}$ and $\hat{\vec{y}}$ components of $\vec{u}$ will always be small, and the equation of motion for $u$ is,
\begin{align}
    \frac{ \partial^2 u }{\partial t^2} & \approx c_\text{l}^2 \frac{\partial^2 u}{\partial z^2} + c_\text{t}^2 \left( \frac{\partial^2 u}{\partial x^2} + \frac{\partial^2 u}{\partial y^2}  \right) \, .
    \label{eq:elastic_wave_equation_isotropic_approx}
\end{align}

We now consider the boundary conditions which must be imposed to find the eigenmode solutions to Eq.~\eqref{eq:elastic_wave_equation_isotropic_approx}. We assume stress-free boundary conditions on each of the BAR surfaces. Since the solutions we are interested in will vanish exponentially near the edge of the BAR, in the $\hat{\vec{x}}$ and $\hat{\vec{y}}$ directions, the only non-trivial boundary condition is on the top and bottom surfaces in the $\hat{\vec{z}}$ direction. Mathematically the stress-free condition is,
\begin{align}
    \hat{\vec{n}} \cdot \bm{\sigma}|_{z = 0} & = \hat{\vec{n}} \cdot \bm{\sigma}|_{z = L(x, y)} = 0 \, ,
    \label{eq:stress_free_exact}
\end{align}
where $\hat{\vec{n}}$ is the vector normal to the surface, $\bm{\sigma}^{ij} = \lambda \delta^{ij}(\nabla \cdot \vec{u}) + \mu (\nabla^i \vec{u}^j + \nabla^j \vec{u}^i)$ is the stress tensor, $\lambda = \rho (c_\text{l}^2 - 2 c_\text{t}^2)$, and $\mu = \rho c_\text{t}^2$. In the limit where the $\hat{\vec{x}}$ and $\hat{\vec{y}}$ components of $\vec{u}$ are negligible ($ \hat{\vec{x}} \cdot \vec{u} \sim \hat{\vec{y}} \cdot \vec{u} \ll u$), $u$ varies slowly in $\hat{\vec{x}}$ and $\hat{\vec{y}}$ ($\partial u / \partial x \sim \partial u / \partial y \ll \partial u / \partial z$), and the top surface position, $L(x, y) = L_0 - \Delta(x, y)$, does not vary rapidly in $\hat{\vec{x}}, \hat{\vec{y}}$ ($\partial \Delta / \partial x \sim \partial \Delta / \partial y \ll 1$) the boundary condition in Eq.~\eqref{eq:stress_free_exact} reduces to,
\begin{align}
    \frac{\partial u}{\partial z}\bigg|_{z = 0} = \frac{\partial u}{\partial z}\bigg|_{z = L(x, y) } \approx 0 \, .
    \label{eq:stress_free_approx}
\end{align}
Even in this approximation, implementing the boundary condition is non-trivial due to the $x, y$ dependence of the top surface position in $\Delta(x, y)$. However assuming $\Delta(x, y)$ is a perturbation, we can change variables,
\begin{align}
    \zeta \equiv \frac{z}{L(x, y)}
\end{align}
and move the $x, y$ dependence in the boundary conditions to the equation of motion such that the system is reduced to,
\begin{align}
    \frac{ \partial^2 u }{\partial t^2} & \approx \frac{c_\text{l}^2}{L_0^2} \left( 1 + \frac{2 \Delta}{L_0} \right) \frac{\partial^2 u}{\partial \zeta^2} + c_\text{t}^2 \left( \frac{\partial^2 u}{\partial x^2} + \frac{\partial^2 u}{\partial y^2}  \right)~~~,~~~\frac{\partial u}{\partial \zeta}\bigg|_{\zeta = 0} = \frac{\partial u}{\partial \zeta}\bigg|_{\zeta = 1} = 0 \, .
    \label{eq:system}
\end{align}

We look for solutions to Eq.~\eqref{eq:system} of the form,
\begin{align}
    u_{n} = \cos{( n \pi \, \zeta )} \, T_n(x, y) \, e^{\pm i \omega_{n} \left( 1 + \, \delta \right) t} \, ,
    \label{eq:u_guess}
\end{align}
where $\omega_{n} \equiv c_\text{l} n \pi / L_0$, and $\delta$ is a small, free parameter. Note that these solutions satisfy the boundary condition in Eq.~\eqref{eq:system}. Substituting Eq.~\eqref{eq:u_guess} in to the equation of motion in Eq.~\eqref{eq:system}, and using the plano-convex top surface function, $\Delta(x, y) = h ( x^2 + y^2 ) / R^2$, renders an equation of motion for $T_n(x, y)$,
\begin{align}
    \delta \, T_n = -\frac{c_\text{t}^2}{ 2 \omega_{n}^2 } \left( \frac{\partial^2 T_n}{\partial x^2} + \frac{\partial^2 T_n}{\partial y^2} \right) + \frac{h}{L_0 R^2} \left( x^2 + y^2 \right) \, T_n \, .
\end{align}
This equation of motion is mathematically identical to the Schr\"{o}dinger equation for a particle of mass $m = \omega_{n}^2 / c_t^2$ in a 2D simple harmonic oscillator with natural frequency $\omega = \sqrt{ 2 h / ( m R^2 L_0 )}$. The solutions are indexed by $\vec{n} = \{ n_x, n_y, n \}$,
\begin{align}
    T_{\vec{n}}(x, y) & = \left[ e^{-x^2/(2 r_n^2)} H_{n_x}\left( \frac{x}{r_n} \right) \right] \left[ e^{- y^2/(2 r_n^2)} H_{n_y}\left( \frac{y}{r_n} \right) \right]~~~,~~~\delta_{\vec{n}} = \frac{c_\text{t}}{\omega_n R} \sqrt{ \frac{2 h}{L_0}} \left( n_x + n_y \right) \, ,
\end{align}
where $H_i$ are the Hermite polynomials, $n_x \geq 0$, $n_y \geq 0$, and $r_n$ is given in the main Letter in Eq.~\eqref{eq:r_n}. The total phonon eigenmodes are then,
\begin{align}
    U_{\vec{n}}(\vec{x}) & = \mathcal{N}_{\vec{n}} \, \cos{\left( \frac{n \pi z}{L(x, y)} \right)} \exp{\left( -\frac{x^2 + y^2}{2 r_n^2} \right)} H_{n_x}\left( \frac{x}{r_n} \right)H_{n_y}\left( \frac{y}{r_n} \right)~~~,~~~\mathcal{N}_{\vec{n}} & = \sqrt{ \frac{1}{M_{n} \,  \omega_{\vec{n}}} } \sqrt{ \frac{1}{2^{n_x + n_y} \, n_x! \, n_y! } } 
\end{align}
which have energy $\omega_{\vec{n}} = (n \pi c_\text{l} / L_0) (1 + \delta_{\vec{n}})$, and $U_{\vec{n}}$ are normalized to $\int  U_{\vec{n}}^2 \, \dd^3 \vec{x} = ( 2 \rho \, \omega_{\vec{n}} )^{-1}$ (assuming $r_n \ll R$) such that $U_{\vec{n}}$ corresponds to a single phonon solution. 

The eigenmodes reported in Eq.~\eqref{eq:phonon_mode_functions} of the main Letter correspond to the $n_x = n_y = 0$ solutions. The $n_x = n_y = 0$ modes are the focus since they have the strongest coupling to a long-wavelength dark electric field. To see this explicitly we can compute the on-resonance power in Eq.~\eqref{eq:P_s_res} for any even $n_x, n_y$ ($P_s^\text{res}(n_x, n_y)$) by a simple rescaling,
\begin{align}
    P_{s}^\text{res}(n_x, n_y) & = P_s^\text{res} \times \frac{\mathcal{N}_{\vec{n}}^2}{\mathcal{N}_{n}^2} \times \left(\frac{\left[\displaystyle\int e^{-x^2/(2 r_n^2)} H_{n_x}(x / r_n) \, \dd x \right] \left[ \displaystyle \int e^{-y^2/(2 r_n^2)} H_{n_y}(y / r_n) \, \dd y \right] }{\left[ \displaystyle \int e^{-x^2/(2 r_n^2)} \, \dd x \right] \left[ \displaystyle \int e^{-y^2/(2 r_n^2)} \, \dd y \right]} \right)^2 \nonumber \\ 
    & \approx P_s^\text{res} \times \frac{1}{ 2^{n_x + n_y} \, n_x! \, n_y!} \times \left( \frac{\left[\displaystyle\int e^{-x^2/(2 r_n^2)} H_{n_x}(x / r_n) \, \dd x \right] \left[ \displaystyle \int e^{-y^2/(2 r_n^2)} H_{n_y}(y / r_n) \, \dd y \right] }{2 \pi r_n^2} \right)^2 \nonumber \\ 
    & = P_s^\text{res} \times \frac{1}{ 2^{n_x + n_y} \, n_x! \, n_y!} \times \left( \frac{n_x! \, n_y!}{(n_x/2)! \, (n_y/2)!} \right)^2 \, , 
\end{align}
which behaves as $P_{s}^\text{res}(n_x, n_y) \sim P_s^\text{res} \times ( n_x n_y )^{-1/2} $ for large $n_x, n_y$.

\end{document}